\documentclass[twocolumn]{aastex62}
\usepackage{xfrac}
\usepackage{amsmath}

\submitjournal{ApJ}

\shorttitle{Flux Power Spectrum}
\shortauthors{Mishra and Gnedin}

\def\dim#1{{\rm #1}}

\begin{document}
\title{Cosmic Reionization on Computers: Evolution of the Flux Power Spectrum}

\correspondingauthor{Nickolay Y.\ Gnedin}
\email{gnedin@fnal.gov}

\author{Nishant Mishra}
\affiliation{Department of Physics; 
The University of California; 
Berkeley, CA 94720, USA}
\affiliation{Fermi National Accelerator Laboratory;
Batavia, IL 60510, USA}
\affiliation{Lawrence Berkeley National Laboratory;
Berkeley, CA 94720, USA}

\author{Nickolay Y.\ Gnedin}
\affiliation{Fermi National Accelerator Laboratory;
Batavia, IL 60510, USA}
\affiliation{Kavli Institute for Cosmological Physics;
The University of Chicago;
Chicago, IL 60637 USA}
\affiliation{Department of Astronomy \& Astrophysics; 
The University of Chicago; 
Chicago, IL 60637 USA}

\begin{abstract}
We explore the evolution of the flux power spectrum in the Cosmic Reionization On Computers (CROC) simulations. We find that, contrary to some previous studies, the shape of the flux power spectrum is rather insensitive to the timing of reionization. However, the amplitude of the flux power spectrum does strongly evolve with time, and that evolution is almost perfectly correlated with the timing of reionization. We show how such correlation can be used in a (futuristic) measurement to determine the redshift of overlap of ionized bubbles.
\end{abstract}


\section{Introduction}
\label{sec:intro}

The Epoch of Reionization (EoR) is a critical period in the thermal history of our universe. This period is when radiation from the first light sources reionized the neutral hydrogen. The Lyman-$\alpha$ (Ly$\alpha$) forest, created by the resonant scattering of redshifted Ly$\alpha$ photons from distant quasars by neutral hydrogen along a line-of-sight, has been used to successfully probe this period. While on large scales the Ly$\alpha$ forest can be well described by the large-scale structure in the $\Lambda$-CDM cosmological model, on small scales it is dependent on the thermal history of the intergalactic medium (IGM). This allows it to be used as a probe of the EoR, as well as other astrophysical phenomena, such as the thermal nature of dark matter. In this paper we utilize the Cosmic Reionization on Computers (CROC) simulation suite to explore the evolution of the one of the key statistical measures of the Ly$\alpha$ forest, its flux power spectrum \citep{Croft2002}, at redshifts $5<z<7$.

Temperature fluctuations affect the Ly$\alpha$ flux power spectrum at a variety of scales. On small scales, Doppler broadening caused by thermal velocities of the neutral hydrogen along the line-of-sight results in the cutoff in the flux power spectrum at $k \gtrsim 0.1\dim{km}^{-1}\dim{s}$. There is also an additional effect of pressure smoothing that contributes to the small scale cutoff and is dependent on the full prior thermal history \citep{Gnedin1998}. At larger scales, inhomogeneous distribution of ionized bubbles during EoR can affect the power spectrum by causing large-scale temperature fluctuations \citep{Onorbe2017,Wu2019,MonteroCamacho2019,MonteroCamacho2020,MonteroCamacho2021,Molaro2021}. Finally, the flux power spectrum can also be used to probe the overall ionized fraction of the universe as function of redshift, since at some sufficiently high redshift during EoR all Ly$\alpha$ forest should become saturated and transmitted flux vanish.

A number of measurements of the flux power spectrum at $z>5$, approaching the epoch of reionization, currently exist \citep{Viel2013,Walther2018,DAloisio2019,Boera2019,Walther2019}. As the observational data on high redshift quasars (and, hence, on the high redshift Ly$\alpha$ forest) are bound to explode with the the next generation of optical telescopes, it is useful to consider what potential constraints can be obtained from the observations of the flux power spectrum at $z\sim6$.

\section{Methodology}
\label{sec:metho}

As a plausible model of cosmic reionization we use simulations from the "Cosmic Reionization On Computers" (CROC) project \citep{gnedin14,gnedinandkaurov_14}. CROC simulations have two properties that make then highly suitable for this project: first, they provide a reasonable, albeit not perfect, match to the observed distributions of optical depths and dark gaps in the Ly$\alpha$ forest at $5\lesssim z \lesssim6$ \citep{gnedin_etal17}. Second, CROC simulations include several random realizations with variable DC modes \citep{gnedin_etal11} for each box size, and these random realization can be used both as samples of different spatial locations in the universe (the primary, designed usage) and as models of different universes with somewhat different reionization histories (a secondary, serendipitous usage). It is the latter usage that is of a particular value for this work.

\begin{figure}[t]
\includegraphics[width=\hsize]{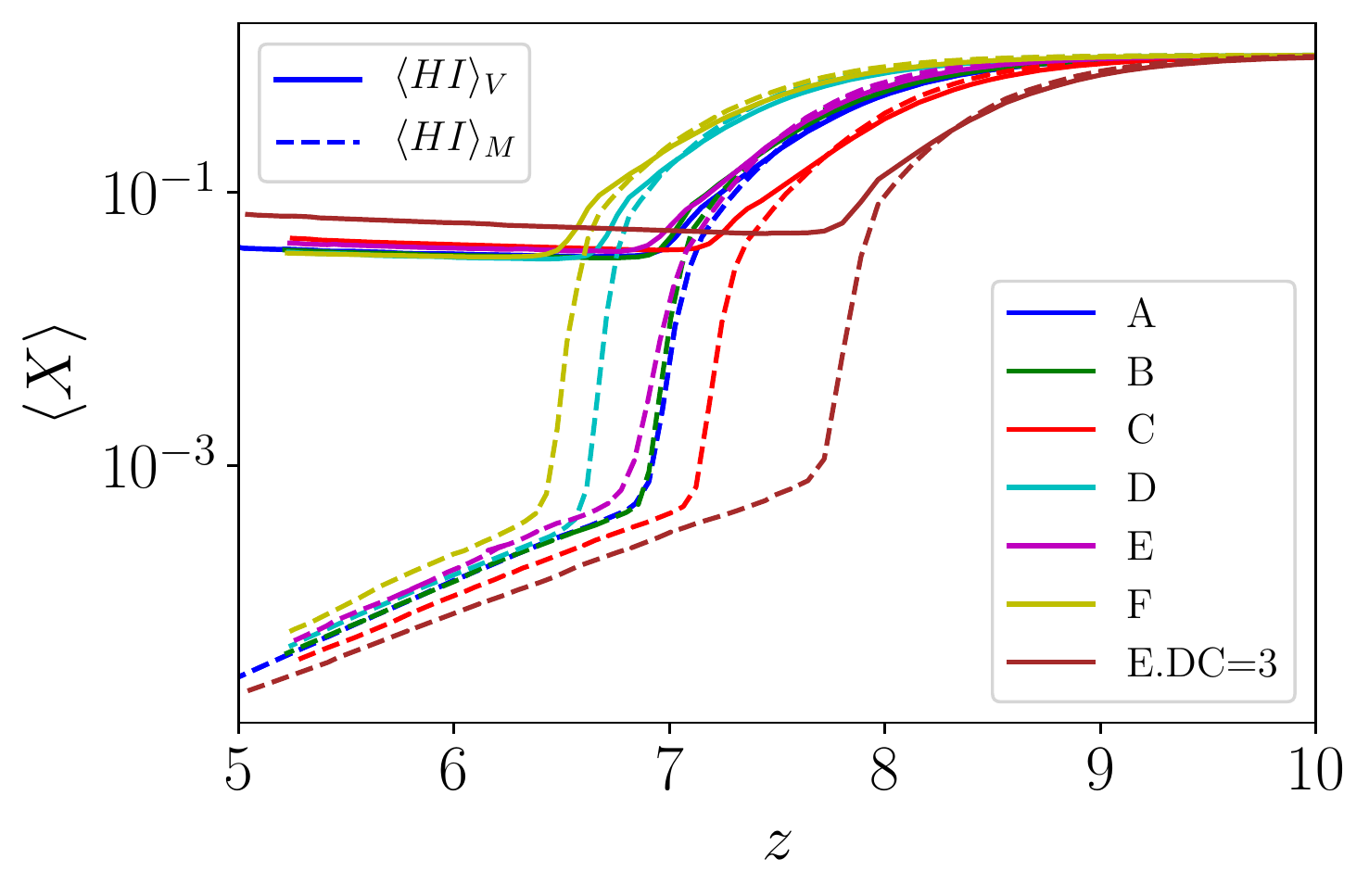}
\caption{Reionization histories for seven independent realizations of the $40h^{-1}\dim{cMpc}$ simulation volume used in this paper.} 
\label{fig:xhz}
\end{figure}

Specifically, we use 7 realizations of $40h^{-1}\dim{cMpc}\approx 60 \dim{cMpc}$ simulation volumes. The largest CROC simulations were performed in $80h^{-1}\dim{cMpc}$ volumes, but volumes of such size are so close to the mean universe that variations in the DC mode do not lead to interesting variations in the reionization histories. In smaller $40h^{-1}\dim{cMpc}$ volumes such variations are larger, and hence these boxes are more useful for our purposes. The 7 realizations include 6 random realizations (that we label A-F) and one variation of box E with the value of the DC mode manually set to $3\sigma$ (to produce a strongly overdense volume that can still be reasonably evolved to $z\sim5$). We label that simulation as "E.DC=3". Figure \ref{fig:xhz} shows reionization histories for these 7 realizations.

For each simulation volume 1000 genuinely random (i.e.\ starting at random positions and going along a random direction) lines of sight were produced at a number of redshifts and Ly$\alpha$ forest spectra were produced for each of them. The (one-dimensional) flux power spectrum $P_F(k)$ was then computed at each redshift as 
\begin{equation}
    P_F(k) = \left\langle \left|\frac{F_k}{\bar{F}}\right|^2 \right\rangle, 
\end{equation}
where $F_k$ is the Fourier transform of the simulated transmitted flux, $\bar{F}$ is the mean flux at that redshift, and averaging is done over all 1000 lines of sight.

\section{Results}

\begin{figure}[t]
\includegraphics[width=\hsize]{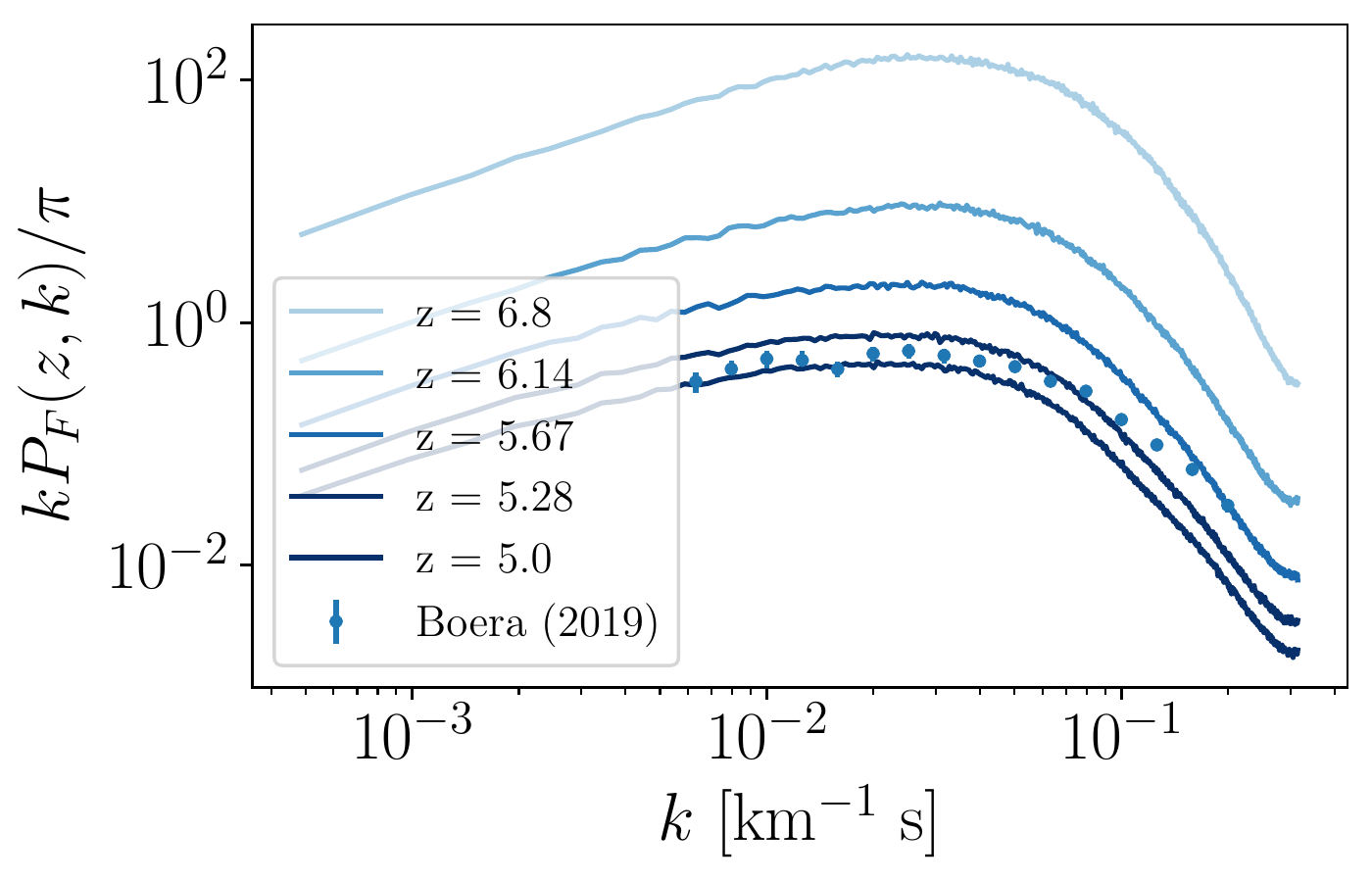}
\caption{The Lyman-$\alpha$ power spectrum from $z=5$ to $z=7$ from realization A. There is higher power at higher redshifts, as regions with transmitted flux become progressively more biased as the redshift increases. Observational data for the Lyman-$\alpha$ power spectrum at $z=5$ from \citep{Boera2019}} are also plotted for comparison.  
\label{fig:compare-ps}
\end{figure}

The Ly$\alpha$ flux power spectra from the realization A are shown in Figure \ref{fig:compare-ps} for multiple redshifts at $5<z<7$. As can be expected, the flux power spectrum grows with redshift, as the higher neutral gas fraction at higher $z$ results in more absorption, while at low $z$ there is more ionized gas. The highest redshift value we show is $z=6.8$. While there is some transmitted flux in some lines of sights at even higher redshifts, the number of lines-of-sight with \emph{any} transmitted flux absorption across the velocity range is too low. This results in a power spectrum that is entirely determined by only a few transmitted spikes. Hence we only show and analyze redshifts where at least 20\% of lines-of-sight contain some flux.

Comparison with the actual measurements of the flux power spectrum at $z=5$ from \citet{Boera2019} shows that realization A slightly underpredicts the amplitude of the flux power spectrum on large scales and falls down at short scales significantly faster than the data. We discuss these deviations below.

\begin{figure}[t]
\includegraphics[width=\hsize]{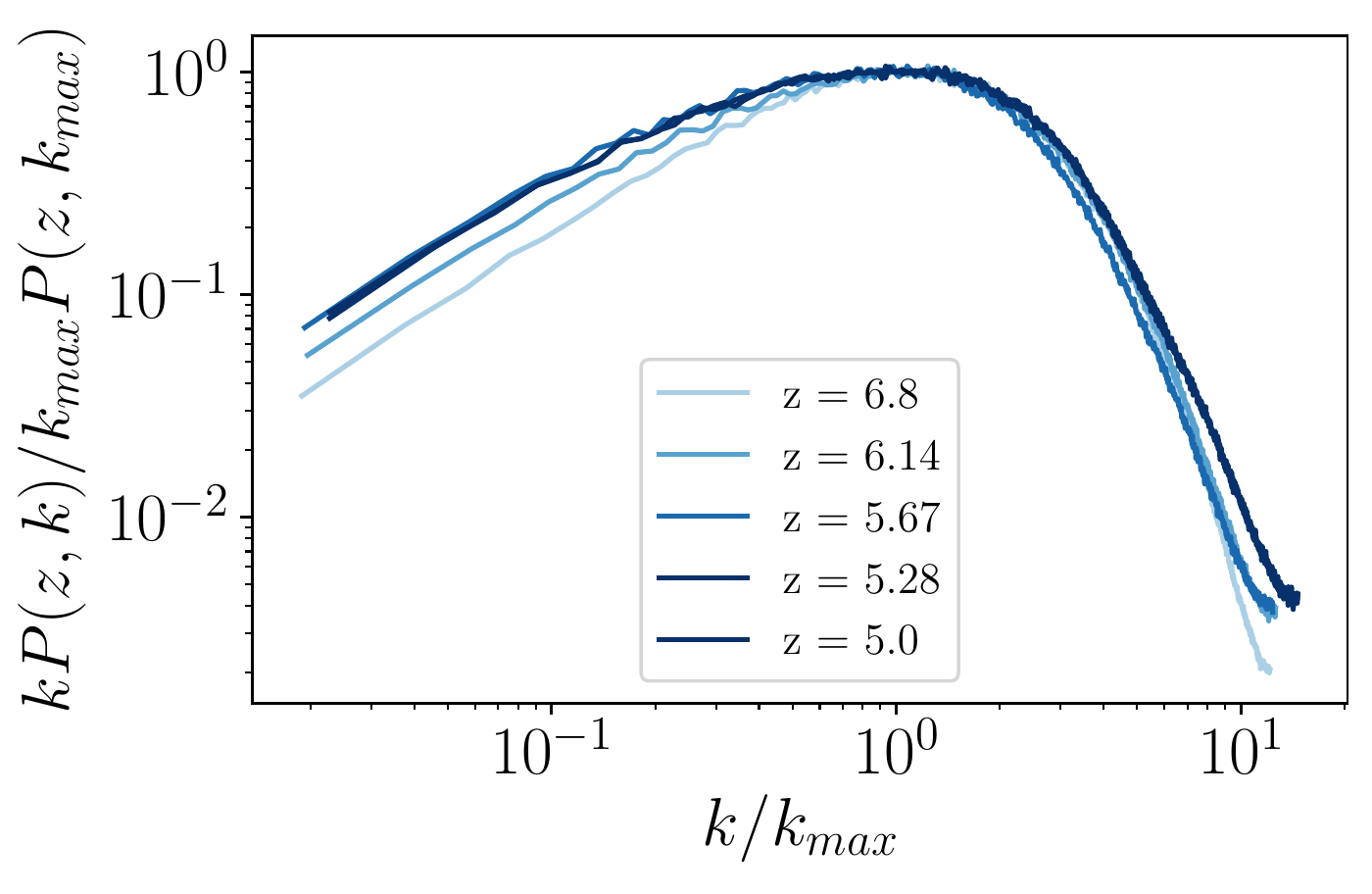}
\caption{Comparison of the relative shape of the Ly$\alpha$ flux power spectrum from $z=5$ to $z=7$. The wavenumbers are scaled by the wavenumber at the maximum $k_{max}$ and vertically scaled by corresponding $P_F(k_{max})$. For a range of $z$, power spectrum shape is highly similar, justifying the use of the peak value of the power spectrum as a summary statistic.} 
\label{fig:shape}
\end{figure}

\begin{figure}[t]
\includegraphics[width=\hsize]{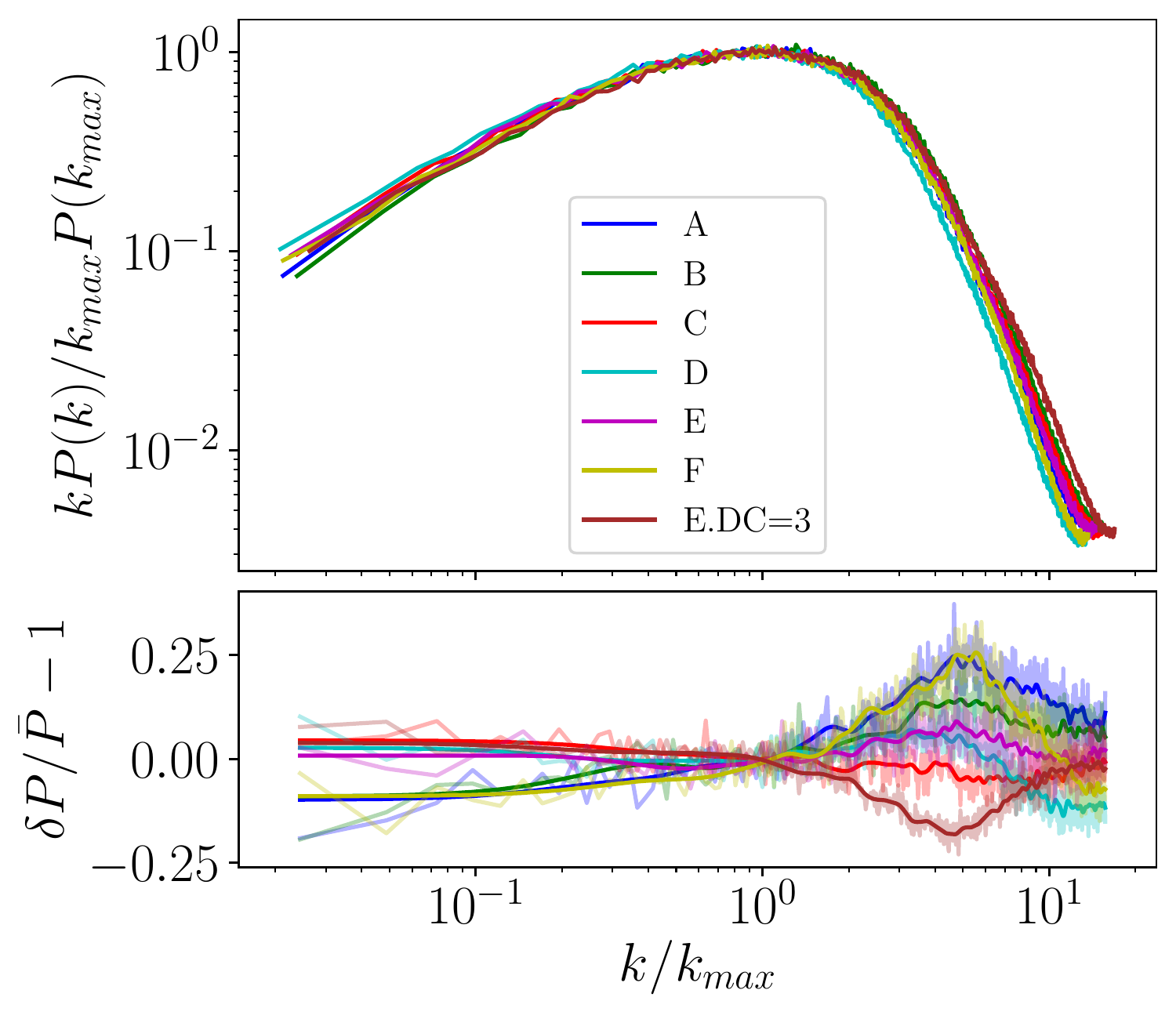}
\caption{Relative shape of the flux power spectrum at $z=5.5$ for each of the 7 realizations. The bottom panel shows the deviations for each of the 7 realization from the best estimate of the cosmic mean (the average of random realizations A, B, C, D, E, F, excluding the biased E.DC=3 one); thin lines show the actual deviations in each wavenumber, solid lines are running averages in 5 bins. There is mild variation in the shape at high and low $k$, but it is probably too small to be of interest.} 
\label{fig:shape2}
\end{figure}

It is clear from Fig.\ \ref{fig:compare-ps} that the shape of the flux power spectrum does not change significantly, while the amplitude evolves rapidly with redshift. In order to illustrate the evolution in the shape (or lack thereof) quantitatively, 
we show in Figure \ref{fig:shape} the flux power spectrum normalized by its value at the maximum. In order to measure the maximum value, the flux power spectrum is smoothed by a 5 pixel wide Gaussian filter to avoid measuring a noisier part of the spectrum as the true $k_{max}$.

The overall shape of the flux power spectrum is consistent across redshifts with only mild variation at the highest redshifts (and even that may be a statistical fluke due to the limited number of transmitted spikes). Figure \ref{fig:shape2} shows the normalized shape for the flux power spectrum for all 7 realization we consider at a fixed redshift. Again, little variation in the shape is observed, although there exists a mild trend of more power at high $k$ for realization with larger DC modes.

In other words, the shape of the flux power spectrum is similar enough across redshifts and realizations that its peak value of $P_F(k_{\rm max})$ is sufficient for analysis of data from any conceivable survey in the near future. Figure \ref{fig:compare-ps} shows that the power at $z \sim 7$ is nearly four orders of magnitude lower than power at $z \sim 5$, and the magnitude at high $k$ and high $z$ is lower still. Given a limited number of quasars available at high redshift, it is unlikely that future surveys will reach the level of sensitivity needed to sufficiently explore the small variations in the shape of the power spectrum we observe in Figs.\ \ref{fig:shape} and \ref{fig:shape2}.

\begin{figure}[t]
\includegraphics[width=\hsize]{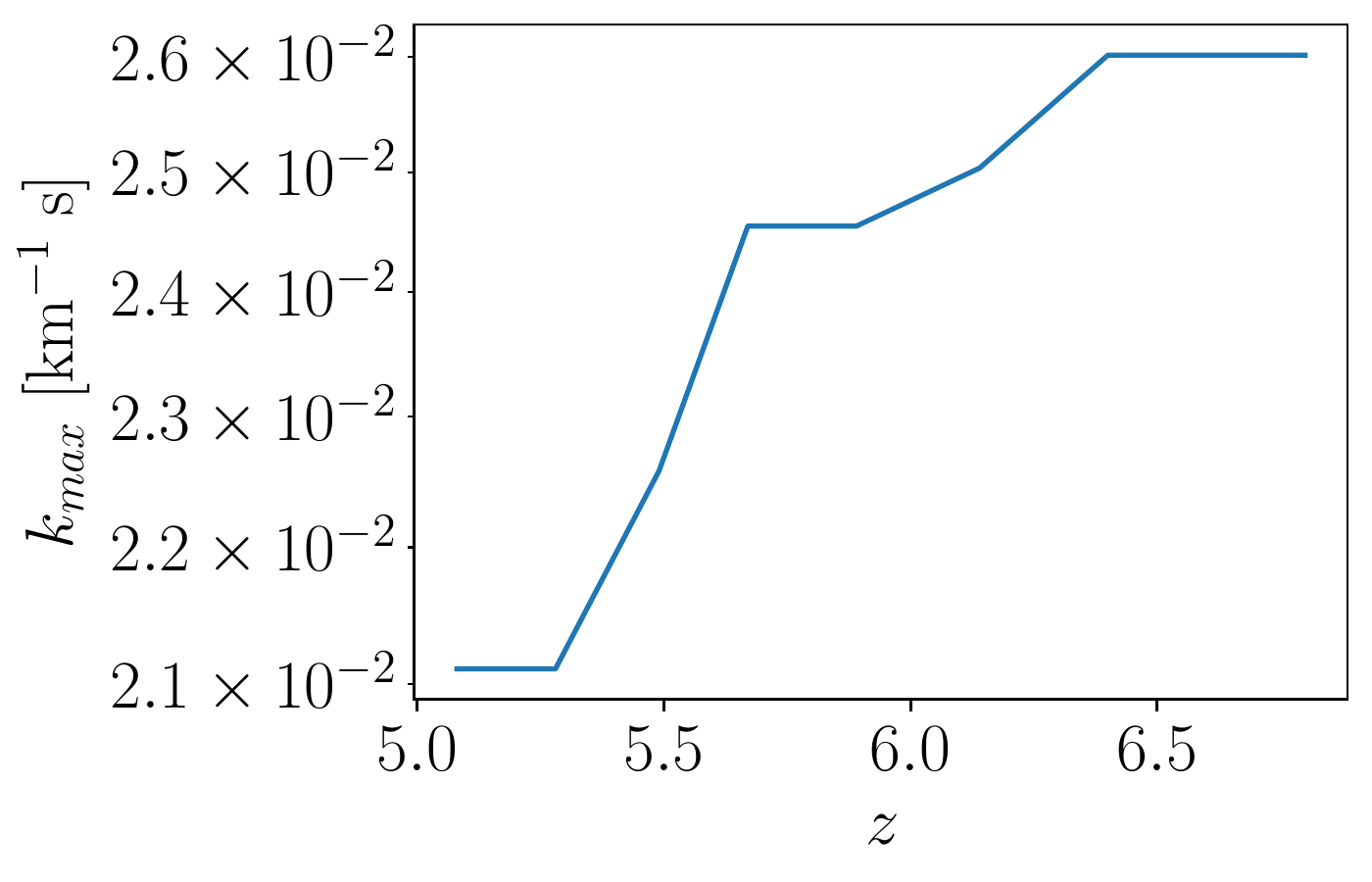}
\caption{Evolution of the wavenumber of the peak of the flux power spectrum $k_{max}$ in realization A. The $k_{\rm max}$ value varies little as a function of redshift.}
\label{fig:kmax}
\end{figure}

If the full flux power spectrum can be characterized by only two values, $k_{\rm max}$ and $P_F(k_{\rm max})$, we only need to explore the evolution of these two quantities. The evolution of $k_{\rm max}$ for realization A is shown in Figure \ref{fig:kmax}. It appears that $k_{\rm max}$ also varies little with redshift, and given the difficulty in measuring the exact location of such a broad maximum, we do not consider $k_{\rm max}$ as a useful probe of reionization history hereafter. That leaves the amplitude of the flux power spectrum $P_F(k_{\rm max})$ as the only quantity of interest.

\begin{figure*}[t]
\includegraphics[width=0.5\hsize]{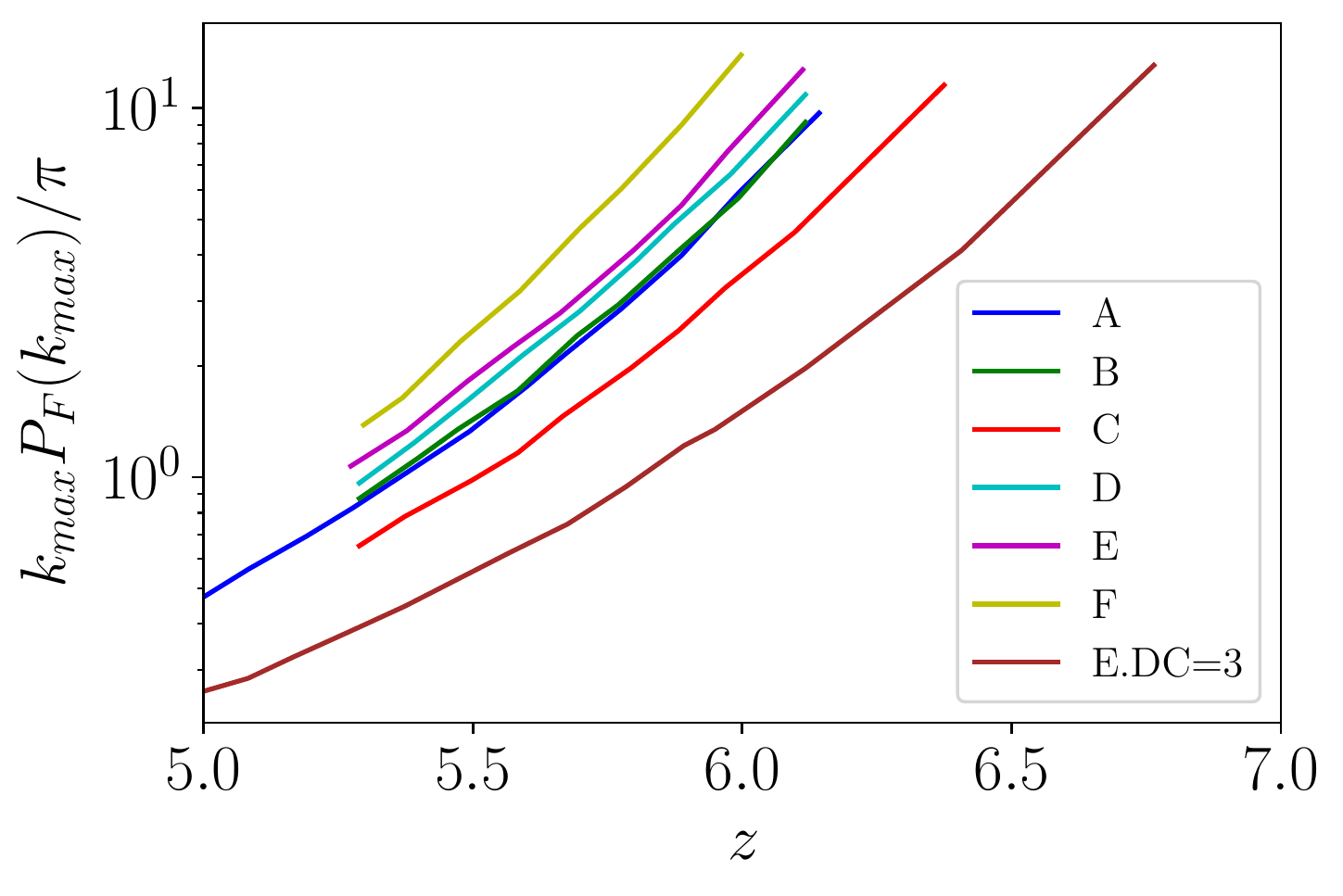}%
\includegraphics[width=0.5\hsize]{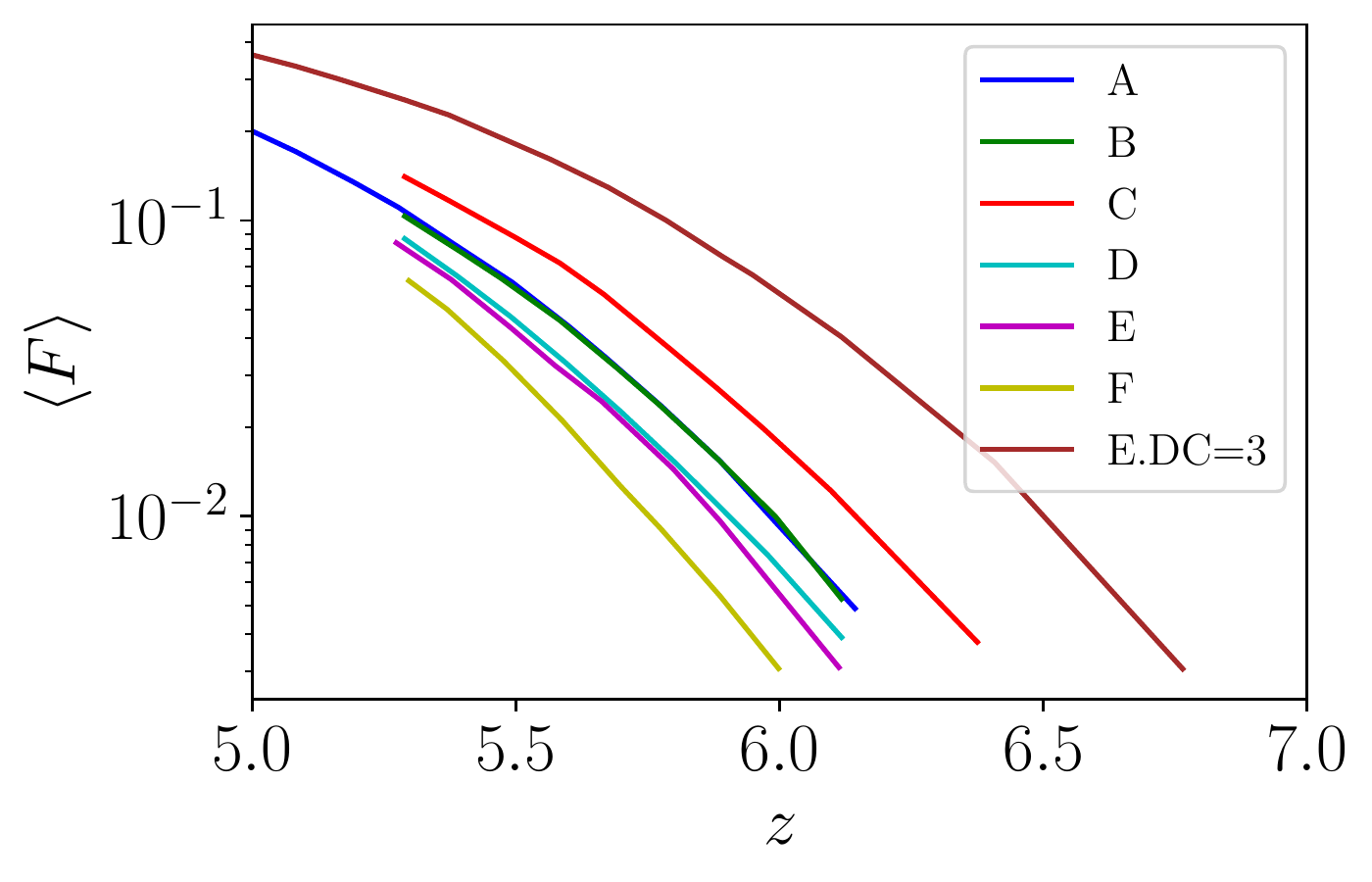}
\caption{Evolution of the amplitude of the flux power spectrum $P_F(k_{\rm kmax})$ (left) and the mean transmitted flux (right) for the 7 independent realizations. The two quantities evolve in opposite directions, but both correlate strongly with the reionization history.}
\label{fig:peaks}
\end{figure*}

The redshift evolution of $k_{\rm max} P_F(k_{\rm max})/\pi$ is shown in the left panel of Figure \ref{fig:peaks} for all 7 independent realizations described in \S\ref{sec:metho}. While the evolution is similar for all 7 simulation boxes, they do exhibit systematic offsets. Comparison with Fig.\ \ref{fig:xhz} shows that models that reionize later reach the same level of flux clustering at later times. We return to this trend below.

The rapid evolution in the amplitude of the flux power spectrum apparent from Fig.\ \ref{fig:compare-ps} is, at least in part, due to the rapid evolution of the overall level of Ly$\alpha$ transmission in the quasar spectra. The evolution of mean transmitted flux $\bar{F}$ for our 7 independent realization is shown in the right panel of Fig.\ \ref{fig:peaks}, and resembles, up to the flip in the direction, the corresponding evolution of the flux power spectrum from the left panel of that figure. 

It is curious that the rate of the evolution in two quantities is similar (up to the opposite direction). That implies that the quantity $\bar{F} P_F(k_{\rm max}) = \left\langle \left|F_k\right|^2/\bar{F} \right\rangle$
evolves little with redshift. We explored this quantity but found no obvious application for its use.

\begin{figure}[t]
\includegraphics[width=\hsize]{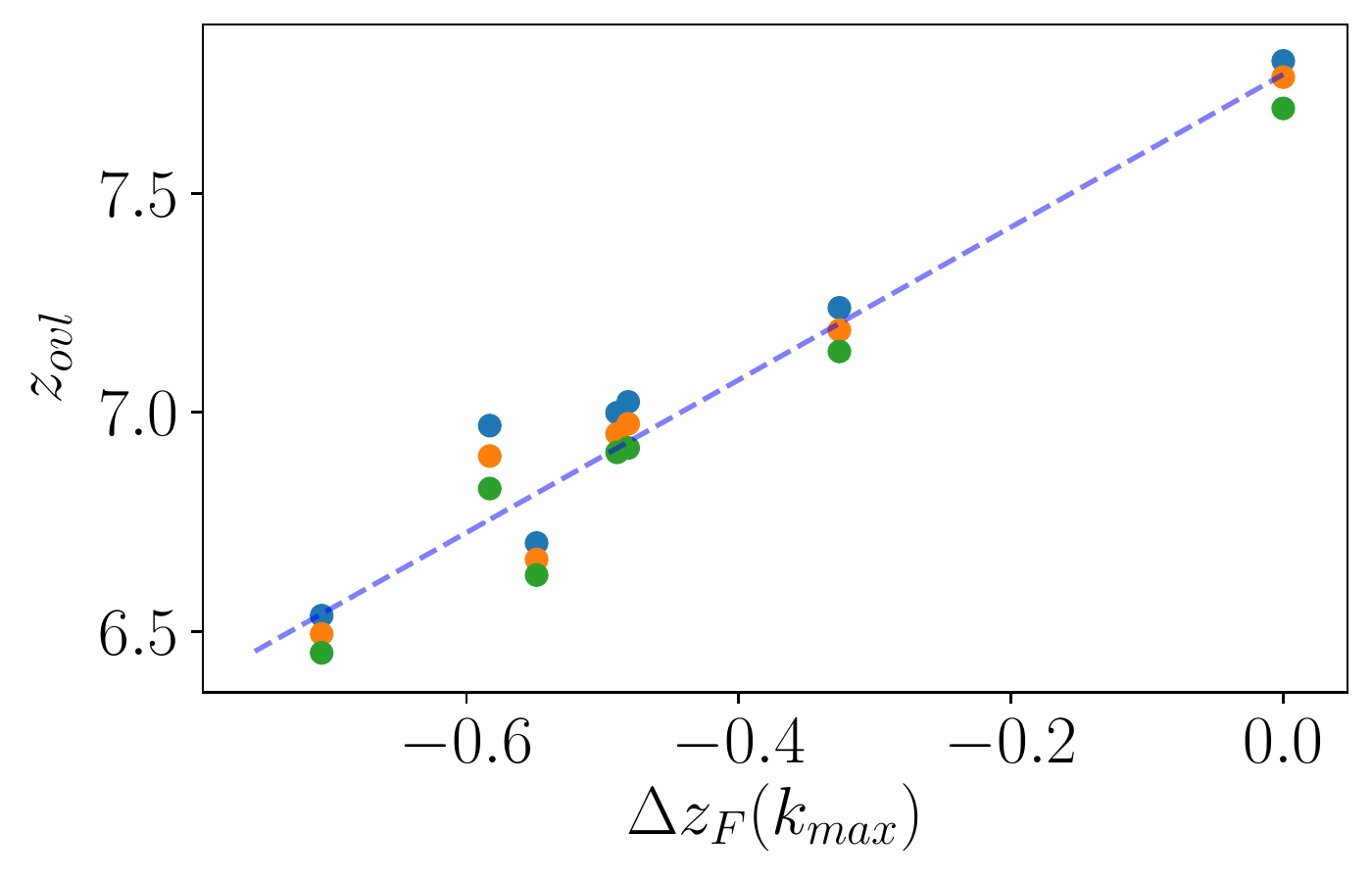}
\caption{Relationship between the redshift offset between different independent realizations (relative to the "E.DC=3" realization) and their corresponding redshift of overlap (defined as the redshift at which $\langle X_{\rm HI}\rangle_V=X_{\rm crit}$). Green, orange, and blue symbols are for $X_{\rm crit}=10^{-4}$, $10^{-3.5}$, and $10^{-3}$ respectively; the observed trend is robust with respect to the definition of the redshift of overlap. The line is the linear best-fit to the relation.}
\label{fig:zvsz}
\end{figure}

In order to quantify the redshift evolution of the flux power spectrum, we measure the redshift offset between each of the 7 independent realization and the "E.DC=3" realization (including its own, zero offset, with itself) as the redshift offset that makes the two curves for $k_{\rm max} P_F(k_{\rm max})/\pi$ versus $z$ coincide on average. This redshift offset can then be compared with the "redshift of overlap" for each of the independent realizations. While the formal definition of the redshift of overlap for cosmic reionization is the moment when the photon mean free path grows at the highest rate \citep{Gnedin2000}, that formal definition is not easy to use in practice. A good proxy for it is the moment when the volume-weighted neutral hydrogen fraction falls below $10^{-3}$ or so. In Figure \ref{fig:zvsz} we show the relation between the so-defined redshift offset and the proxy for the redshift of overlap for three different values of the volume-weighted neutral hydrogen fraction.

\section{Conclusions}

In CROC simulations the shape of the flux power spectrum at large scales ($k < 0.005 {\rm km^{-1}s}$) varies little (at the level of 10\%) with the redshift of the measurement (except at highest redshifts) or the redshift of overlap, and that result is in surprising disagreement with the conclusions from AREPO-RT simulations \citet{Wu2019}, who found systematic deviations at large scales of the order of 40\%. The reason for such discrepancy is unclear, given that both sets of simulations have comparable box sizes, mass resolution, and radiation frequency coverage. Both simulations use the moments method for modeling radiative transfer, with the only difference that CROC employs OTVET approximation and AREPO-RT uses the M1 closure. There are no known numerical artifacts in either of the scheme, though, that could explain the difference between the two sets of simulations.

With this disagreement noted, we find that the amplitude of the flux power spectrum (quantified by the value at the maximum, $P_F(k_{\rm max})$), evolves rapidly with redshift. That evolution reflects the rapid change in the ionization state of the IGM at the end of reionization and is similar (but not identical) to the evolution of the averaged transmitted flux, $\langle F\rangle$. 

\begin{figure}[t]
\includegraphics[width=\hsize]{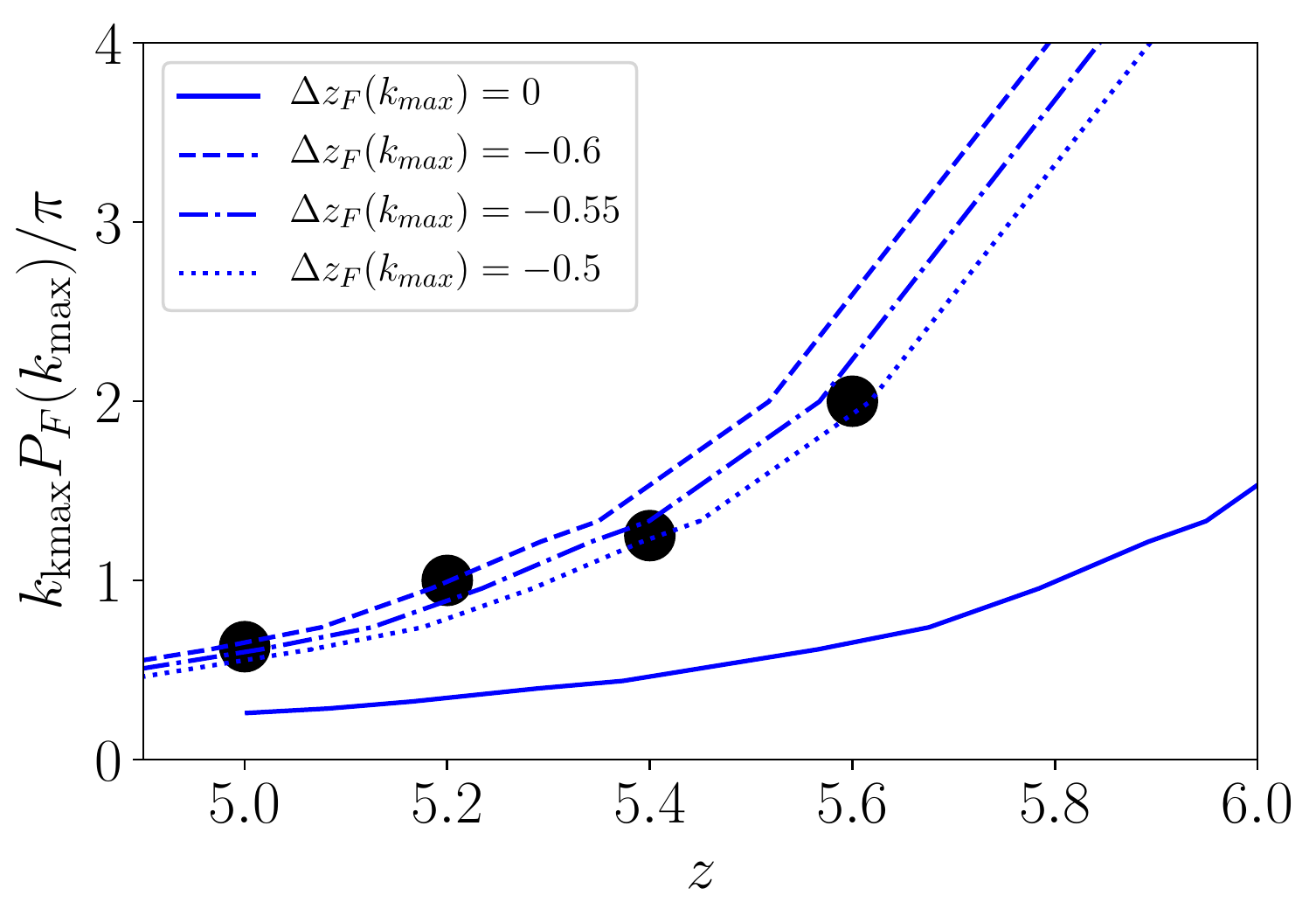}
\caption{Demonstration of the potential method for measuring the redshift of overlap. Estimates for the observed values of $k_{\rm max} P_F(k_{\rm max})/\pi$ from \citet{Boera2019} and \citet{DAloisio2019} are shown with black circles (these are estimates ``by eye" sufficient for this demonstration and are not exact measurements). The solid blue line is the reference E.DC=3 model, while the other three lines are the reference model shifted in redshift by the amount indicated in the legend.}
\label{fig:zoff}
\end{figure}

The most interesting property of that evolution, however, is that it is correlated with the evolution of the mean neutral fraction. The latter, if volume averaged, serves as a good proxy for the redshift of overlap of ionized bubbles. Hence, measuring the rate of evolution of the flux power spectrum at $z\sim5-5.5$ allows one to constraint the redshift of overlap. An example of how this can be done is shown in Figure \ref{fig:zoff}. The best fit to the observational data is obtained by shifting the reference model E.DC=3 by $\Delta z=-0.55$. Comparison with Fig.\ \ref{fig:zvsz} shows that such a redshift shift corresponds to the redshift of overlap of about 6.7.

This measurement relies on the calibration from the simulations, and hence needs high fidelity simulations. CROC simulations can serve as an illustration of this procedure, but are of not high enough fidelity to be used for the actual measurement - they provide a close match to the observed distributions of opacities in the IGM and sizes of dark gaps \citep{gnedin_etal17}, but not close enough to match all the observational data within the observational uncertainties. Hence, we only present Fig.\ \ref{fig:zoff} as an illustration. 

The expectation that simulations become sufficiently good to allow such a measurement is not futuristic, though. CROC simulations do not provide the exact match to the data, but they are off by a factor of 2 at most, so the next generation simulations (circa 2025) can be expected to provide an order of magnitude improvement to the current ones and offer a sufficiently accurate match to only only existing but future observational data as well.

\acknowledgments
This manuscript has been authored by Fermi Research Alliance, LLC under Contract No. DE-AC02-07CH11359 with the U.S. Department of Energy, Office of Science, Office of High Energy Physics. This work used resources of the Argonne Leadership Computing Facility, which is a DOE Office of Science User Facility supported under Contract DE-AC02-06CH11357. An award of computer time was provided by the Innovative and Novel Computational Impact on Theory and Experiment (INCITE) program. This research is also part of the Blue Waters sustained-petascale computing project, which is supported by the National Science Foundation (awards OCI-0725070 and ACI-1238993) and the state of Illinois. Blue Waters is a joint effort of the University of Illinois at Urbana-Champaign and its National Center for Supercomputing Applications. This work was supported in part by the U.S. Department of Energy, Office of Science, Office of Workforce Development for Teachers and Scientists (WDTS) under the Science Undergraduate Laboratory Internships (SULI) program.

\bibliographystyle{aasjournal}
\bibliography{main}

\end{document}